\documentstyle[12pt]{article}

\begin{document}
\bibliographystyle{unsrt}

\begin{flushright} UMD-PP-95-61

November,1994
\end{flushright}

\vspace{6mm}

\begin{center}

{\Large \bf Almost Degenerate Neutrinos with Maximal Mixing}\\ [6mm]

\vspace{6mm}
{\bf{R.N. Mohapatra\footnote{Work supported by the
 National Science Foundation Grant PHY-9119745}~and~
S. Nussinov\footnote{Permanent address: Department of Physics and
Astronomy, Tel Aviv University, Tel Aviv, Israel.}}}

{\it{ Department of Physics and Astronomy,}}
{\it{University of Maryland,}}
{\it{ College Park, MD 20742 }}

\end{center}
\vspace{20mm}
\begin{center}
{\bf Abstract}
\end{center}
\vspace{6mm}

We point out that if
the observed deficit of solar and  atmospheric neutrinos are to be understood
as consequences of oscillations between different neutrino flavors,
the simplest way to reconcile it with mixed dark
matter picture of the universe and a possible sub-electron volt
upper limit on the $\nu_e$ Majorana mass is to have
a  scenario of three  light almost {\it degenerate}
Majorana neutrinos ($\nu_e$, $\nu_{\mu}$ and $\nu_{\tau}$ )
 with {\it  maximal} CP-violating mixing among
 among all three of them. We discuss
theoretical scenarios which may lead to such a mass and mixing pattern.
We also discuss a scenario where two of the three degenerate neutrinos
are maximally mixed.

\newpage

\noindent{I.} The conventional  hierarchical
"see-saw" mechanism\cite{seesaw} for
neutrino masses and mixings provides an overall view of neutrino physics
which is quite attractive. In this approach, the left-handed
neutrinos are Majorana particles and their masses are given by:
$$m_{\nu^L_{i}}\simeq {({m^D_{\nu_i}})^2/{M_{N_R}}}\;\eqno(1)$$

where the $m^D_{\nu_i}$ is the Dirac mass connecting the active
left-handed flavor of the neutrino $\nu_{iL}$ with the sterile right-handed
one denoted by $N_{iR}$ and $M_{N_R}$ is the  Majorana mass
for the heavy sterile right-handed neutrino. The formula in equation (1)
is obtained by diagonalizing a generic  $\nu_L$-$N_R$ mass matrix of
the form:
$$M=\left(\begin{array}{cc}
0  &  m^D_{\nu}\\
{m^D_\nu}^T & M_{N_R} \end{array} \right)   \eqno(2)$$

Let us note the particular feature of this matrix that the $\nu_L \nu_L$
entry is assumed to vanish. The light neutrino masses then exhibit the
hierarchical mass pattern if the Dirac masses are assumed to be related
to the charged fermion masses of the corresponding generation and
no strong hieararchy is assumed to exist among the right-handed neutrino
masses. In grand
unified theories such as SO(10) or $E_6$, such relationsbetween the
Dirac masses of neutrinos and the charged fermion masses do indeed arise.
Likewise one expects also a hierarchical pattern in the mixings between
the different generations. Detailed predictive grand unified models
based on the $SO(10)$ group do indeed confirm these intuitive expectations
 for the hierarchical pattern. As far as the absolute values
of the masses is concerned, it is clearly determined by the magnitude
of $M_{N_R}$ which is proportinal to the breaking of the local $B-L$
symmetry in the theory. In the framework of grand unified theories such
as SO(10), the present LEP data completely determine the value of the
$B-L$ symmetry breaking scale , $M_{BL}$. For non-supersymmetric models
\cite{chang}, this scale is anywhere between $10^{10}$GeV to $10^{13.5}$
GeV and for the supersymmetric case , it is around $10^{15}$ to
$10^{16}$GeV. The conventional
see-saw formula then implies that the value of the electron neutrino mass
is in the micro-eV range, that of muon neutrino is in the milli-eV range
and that of the tau neutrino is in the eV range . This kind of neutrino
spectrum has the following experimental implications; (i) it can account
for the hot dark matter (HDM) of the universe if $m_{\nu_{\tau}}\simeq 5-7$ eV;
(ii) it can explain all four solar neutrino experiments using the
attractive MSW mechanism for $\nu_e-\nu_{\mu}$ oscillation for $m_{\nu_{\mu}}
\simeq 10^{-3}$ eV as expected from these models. However given a scenario
that explains the solar neutrino puzzle and HDM of the universe, it cannot
explain the deficit of atmospheric muon neutrinos. This picture can be
confirmed by the proposed CHORUS\cite{ch} and NOMAD\cite{ch}, implying
some new mechanism to understand the atmospheric neutrino deficit other
than neutrino oscillations.

 It was proposed
in ref.\cite{cald}, that the simplest way to accomodate the solar neutrino
puzzle, the atmospheric neutrino puzzle and the MDM picture of the universe
is to assume that the three known neutrinos are almost degenerate with
a common mass of around 2 eV. The reason for near degeneracy is that
the mass difference between $\nu_{\mu}$ and $\nu_{\tau}$ required to
solve the atmospheric neutrino deficit is\cite{gaisser} of order
$5\times 10^{-3}$ to $.5$ $eV^2$. So if we want a $\nu_{\tau}$ with a
mass in the few eV range to become the hot dark matter, $m_{\nu_{\mu}}$
must be in the eV range and be almost degenerate with the $\nu_{\tau}$.
Now since all solutions of the solar neutrino puzzle using neutrino
oscillations also require that
the $\nu_e-\nu_{\mu}$ mass difference must  be very small (i.e.
$10^{-5}$ $eV^2$ in the case of MSW solution or $10^{-10}~eV^2$ in the
case of vacuum oscillation), we must have $\nu_e$ and $\nu_{\mu}$
nearly degenerate. A complete understanding of all three puzzles
therefore require that the three light neutrinos must be degenerate
with masses around 2 eV or so.

Theoretical appeal for such models owes
its origin to an early observation\cite{mohsen} that in many models that
implement the see-saw mechanism, the actual seesaw mass matrix has the form:
$$M=\left(\begin{array}{cc}
m_{LL} & m^D_{\nu}\\
{m^D_{\nu}}^T & M_{N_R} \end{array} \right) \eqno(3)$$

where $m_{LL}=f{{v_W}^2\over{V_{BL}}}$ and $M_{N_R}=f v_{BL}$. The
coefficient $f$ is matrix in generation space, which is not expected
to have a strong dependence on the generations. Note that
$m_{LL}$ leads to a direct mass for the light neutrinos which
while being of see-saw type (i.e. inversely proportional to $v_{BL}$, can
 be assumed to be independent
of generations and will not therefore have the hierarchical pattern given
in eq.(1).  The correct pattern of neutrino masses given by the
the eq.(3) is then:
$$m_{\nu_i}\simeq f_i {{v_W}^2\over{V_{BL}}}~-~{{m^D_{\nu_i}}^2\over{f_i v_{BL}
}} \eqno(4)$$

Note that if all $f_i$'s
 are set to be equal ( perhaps by some symmetry\cite{many} ),
then we have the desired almost degenerate scenario .

An unavoidable prediction of these models is that $m_{\nu_e}\simeq 2$ eV
or so as mentioned. Since in the usual see-saw picture, the
neutrinos are Majorana fermions,
 the present generation of neutrinoless double beta
decay experiments\cite{klap} should observe a signal at the
appropriate level. It appears  that the
Heidelberg-Moscow double beta decay experiment has now an upper limit
on the $\nu_e$ mass $.68$ eV\cite{klap2} at the $90\%$ confidence level.
A question that now arises is whether the
almost degenerate neutrino mass scenario can be maintained while not
being inconsistent with the sub-eV mass limits on $\nu_e$ suggested
in experiments. One could perhaps assume that there are
 uncertainties in nuclear matrix elements which allow the degenerate
scenario to remain viable.
There are however  recent cosmological observations indicating
a higher value of Hubble constant ( $h\simeq .80$ in units of 100km/sec/Mpc )
which taken seriously, would imply that
 a $30\%$ hot dark matter content in the degenerate
neutrino scenario would require common mass $m_{\nu_i}\simeq 6$ eV.
It would certainly be hard to reconcile such a mass value with
the $\beta\beta_{0\nu}$ results.

In this letter, we suggest that a way to reconcile the possible
sub-eV $\beta\beta_{0\nu}$ upper limit on $m_{\nu_e}$
with the almost degenerate
neutrino spectrum is to assume that the light neutrinos have
a maximal CP-violating mixing among themselves\cite{nuss,wolf}. This
may help to reconcile {\it qualitatively} all known data for the neutrinos.
This mixing pattern is a generalization of the symmetric maximal
mixing scheme proposed in\cite{nuss} in order to understand the
reduction of the neutrino flux in the chlorine experiment. We also
note  an alternative scenario where one has a
maximal mixing between $\nu_e$ and one of the other neutrinos .

\vspace{6mm}

\noindent{\it II. Sub-eV effective $\nu_e$ mass and maximally mixed
neutrinos:}

\vspace{4mm}

The starting point of the present paper is the well known result that
in neutrinoless double beta decay, one measures the effective mass
of the light neutrino
$\langle m_{\nu_e}\rangle \equiv \Sigma_{i} U^2_{ei} m_{\nu_i}$
and therefore $\beta\beta_{0\nu}$ amplitude is not only sensitive to
the absolute values of the neutrino mass but also to the their mixing
pattern. It is then clear that the simplest way to reconcile
the almost degenerate scenario for neutrinos
 with a sub-eV upper limit on the effective
Majorana mass of $\nu_e$  as
given by neutrinoless double beta decay experiments is that the neutrino
mixing matrix  be maximal and CP-violating among all three or between
any two generations one of which is the $\nu_e$. We will discuss both
these possibilities below separately.
In what follows , we define the neutrino mixing
matrix as follows: $L_{wk}~=~{{g}\over{2\sqrt{2}}}W^-\bar{e}_{iL}U_{\nu ij}
\nu_{jL}~+~h.c.$ where we have omitted the Lorentz vector character of the
interaction.

\vspace{4mm}
\noindent{\it Case A: Maximal mixing among three generations:}
\vspace{4mm}

The form of the mixing matrix in this case is given by:

$$U_{\nu}={{1}\over{\sqrt{3}}}\left(\begin{array}{ccc}

1 & \omega & {\omega}^2 \\
1  & {\omega}^2 & \omega \\
 1 & 1 & 1  \end{array} \right)   \eqno(5)$$
where $\omega=e^{{2\pi i}\over{3}}$. Note that any redefinition
of fields to remove these phases makes the Majorana masses complex
and leads to the same result.
If we write the neutrino masses to be $m_{\nu_e}= m_0 + \delta_e$,
$m_{\nu_\mu}= m_0 +\delta_{\mu}$ and $m_{\nu_\tau}= m_0 + \delta_{\tau}$
with $\delta_e,~\delta_{\mu},~\delta_{\tau}\ll m_0\simeq few~eV$, the
effective mass measured in the $\beta\beta_{0\nu}$ experiment is
given by $\langle m_{\nu_e} \rangle \equiv \Sigma_{i=\nu_e,\nu_{\mu},\nu{\tau}}
 U^2_{ei}~m_{i}~=~ \delta_e +\omega^2 \delta_{\mu}+ \omega \delta_{\tau}$,
( which follows from the fact that $1+\omega+\omega^2=0$)
 which is clearly in the sub-eV range. In fact, the
solar and atmospheric neutrino data imply that $\delta_e,\delta_{\mu}
\ll \delta_{\tau}\leq .25~eV$ (or $.08~eV$ )
if we assume, $m_0=2~eV$ ( or 6 eV). Thus, if the
atmospheric neutrino data becomes precise enough to fix the $\Delta^2
_{\mu-\tau}$, the $\beta\beta_{0\nu}$ decay experiment
can be used to test this hypothesis since both the Heidelberg-Moscow
$^{76}Ge$ experiment and the NEMO 3 experiment using $^{100}Mo$\cite{nemo}
can probe effective Majorana $\nu_e$ mass
$\langle m_{\nu_e} \rangle$ down to $.1$ eV.

Let us now briefly address the question of how one understands the
solar and atmospheric neutrino data in this picture. The popular
MSW explanation of the solar neutrino puzzle is not applicable in
this case due to large values of the mixing angles\cite{krastev}.
  There are however two other possible approaches using vacuum oscillation.
 It was noted sometime back\cite{nuss}
that if the $\nu_e-\nu_{\mu}$ mass difference square is much larger
than $10^{-10}$ $eV^2$, then the solar $\nu_e$ flux gets suppressed
by a factor $1/3$. While this is perfect for understanding the Chlorine
data, its predictions for the Kamiokande and Gallium data\cite{solar}
are too small compared to observations. The other possibility has
been discussed in a recent paper by Kim and Lee\cite{kim} where one makes the
choice of mass differences $\Delta m^2_{e\mu}\simeq 10^{-10}~eV^2$ and
$\Delta m^2_{\mu-\tau}\simeq 10^{-1}-10^{-3}~eV^2$.
As emphasized in ref\cite{kim},
in the case of the atmospheric neutrinos, one has both the $\nu_e-\nu_\mu$
as well as $\nu_{\mu}-\nu_{\tau}$ oscillation operative and one can fit the
atmospheric neutrino data. As far as the solar neutrino data is concerned,
it is not easy to fit all data simultaneously\cite{kim}; however, allowing
for two standard deviations in the gallium data, a fit has been obtained
in ref.\cite{kim} . It therefore follows that if present data in all solar
neutrino experiments are confirmed by future experiments, this maximal
mixing scenario will
be ruled out. At the present time, however, entertaining such a possibility
is perhaps reasonable.

\vspace{4mm}
\noindent{\it Case B: Maximal mixing between $\nu_e$ and $\nu_{\tau}$:}
\vspace{4mm}

There are two possibilities for mixing in this case: either between $\nu_e$-
$\nu_{\mu}$ or between $\nu_e$ and $\nu_{\tau}$. Let us illustrate
this using the
second case. The neutrino mixing matrix then takes the following form:

$$U_{\nu}=\left(\begin{array}{ccc}
{{1}\over{\sqrt{2}}} & {{\beta}\over{\sqrt{2}}} & {{i}\over{\sqrt{2}}}\\
-\beta c~+{{is}\over{\sqrt{2}}} & c~+{{is\beta}\over{\sqrt{2}}} &
 {{s}\over{\sqrt{2}}}\\
\beta~s~+{{ic}\over{\sqrt{2}}}&-s~+~{{i\beta c}\over{\sqrt{2}}}& {{c}\over
{\sqrt{2}}}  \end{array}    \right)   \eqno(6)$$

We assume that $\beta\ll 1$ and $c$ and $s$ stand for the sine and cosine
of an angle which will be fixed by the atmospheric and solar neutrino data.
To see if this pattern is viable, let us define $\Delta_{ij}=|m^2_{\nu_i}-
m^2_{\nu_j}|$ and $y_{ij}(L)\equiv sin^2\left({{\Delta_{ij}L}
\over{4E}}\right)$. There are
then two cases: {\it (i)}: $\Delta_{12}\ll \Delta_{13}\simeq \Delta_{23}$,
which corresponds to $\delta_e,\delta_{\mu}\ll \delta_{\tau}$;
{\it (ii)}: $\Delta_{13}\ll \Delta_{12}\simeq\Delta_{32}$ corresponding to
$\delta_e,~\delta_{\tau}\ll \delta_{\mu}$.

In case (i),  the survival probability for the solar neutrino
in the vacuum oscillation approximation, is given by:
$$P(\nu_e\rightarrow \nu_e)\simeq 1-~y_{13}(L_S)\eqno(7a)$$
where $L_S$ denotes the Earth-Sun distance.
For the atmospheric neutrinos, we get in the simple approximation
where we ignore the effects of flux and effects at the detector,
$${{N_{\mu}}\over{N_e}}\simeq 2 {{1-(c^2s^2+.5s^2)y_{13}(L_A)}
\over{1-(1-2s^2)y_{13}(L_A)}}\eqno(7b)$$
where $L_A$ denotes the typical distance travelled by the atmospheric
neutrinos ( of the order of $10^3$ to $10^4$ Km.).
 It is then clear that for $y_{13}(L_A)\approx 1$, there exists
a range of values for the free mixing parameter $s$ for which one can fit
the atmospheric neutrino observations.
 However, for these parameters,  eq.(7a) for the case of solar
neutrinos gives a prediction that solar
neutrino flux is energy independent and is reduced by $50\%$ in all
experiments ( since $y_{13}\simeq .5$ )
which is in disagreement with observations.

Turning now to case (ii), we get for the reduction in solar neutrino flux,
$$P(\nu_e\rightarrow\nu_e)\simeq 1~-~y_{13}(L_S)\eqno(8a)$$
In order to understand the solar neutrino data, we would like $y_{13}(L_S)$
to be of order one , which requires $\Delta_{13}\simeq 10^{-10}~eV^2$ . A
careful analysis\cite{bere} in this case leads to the conclusion
that while it possible to obtain a fit to the present data, it is not
a very good fit, though it cannot be definitely excluded.
Using the fact that $y_{13}(L_S)\simeq 1$ implies $y_{13}(L_A)\ll 1$ ,
for the case of atmospheric neutrinos, we find :
$${{N_{\mu}}\over{N_e}}\simeq 2 (1~-~4c^2s^2y_{12}(L_A))\eqno(8b)$$
 As far as the atmospheric neutrino data is concerned,
 one can obtain a good fit assuming $\Delta_{12}\simeq 10^{-2}~eV^2$ .
So unlike the case (i) described above, case (ii) may be acceptable as
a viable scenario  until the  solar neutrino data
becomes more definitive.

In summary, we feel that two of the scenarios outlined above may provide
a reasonable description of all existing observations
 on neutrinos ( i.e. solar, atmospheric, $\beta\beta_{0\nu}$ and
mixed dark matter picture of the universe) and will be testable
once the planned solar neutrino as well as $\beta\beta_{0\nu}$ decay
are carried out. To the extent that such an approach to understanding
all available data using only three Majorana neutrinos
was not discussed in literature, this work should be both of theoretical
and experimental interest. It is also worth pointing out here that
there are early results from the LSND experiment\cite{louis},
which seem to indicate the oscillation of $\nu_{\mu}$ to $\nu_e$
with a $\Delta_{e\mu}\simeq 6~eV^2$. Should this result be confirmed
after further data taking, all the scenarios
discussed in this paper will be ruled out. Finally,
of all the above scenarios, the  one with
all three neutrinos maximally mixed is an extremely symmetric possibility.
We would therefore like to see if it can be derived in a plausible
extension of the standard model.

\vspace{4mm}

\noindent{\it III. Possible theoretical understanding of the maximally
mixed degenerate neutrinos:}

\vspace{4mm}

Let us now explore possible theoretical schemes for generating the
maximally mixed three generation scenario, which, among all three
scenarios described above may have the  best
chance of arising from some underlying symmetry. While we
have not succeeded in finding a gauge theory where the above scheme
can be generated in a  technically natural manner, we have found
a symmetry that such a theory ought to have in order to lead both to
an almost degenerate mass pattern for neutrinos and maximal
CP-violating mixing.

We consider the following extension of the standard model where the fermion
sector is augmented by the addition of the three right-handed neutrinos
( denoted by $\nu_{Ri}$ where $i$ is the generation index). Let us only
focus on the lepton sector and denote the $SU(2)_L$ doublets by $\psi_{Li}$
and the right-handed singlets by $\ell_{iR}$. Let us assume that the
theory prior to spontaneous symmetry breaking is invariant under a $Z_3$
symmetry,which has three elements $( 1, S, S^2)$ ( i.e. $S^3=1$ ). Under
the action of $S$, we have $\psi_{1L}\rightarrow \psi_{2L}$; $\psi_{2L}
\rightarrow \psi_{3L}$ and $\psi_{3L}\rightarrow \psi_{1L}$. So in
generation space, we can write $S$ as a matrix:
$$S=\left(\begin{array}{ccc}
0 & 1 & 0\\
0  &  0 & 1 \\
1 & 0 & 0 \end{array} \right) \eqno(9)$$

As for the remaining fermion fields, we assume that under the $Z_3$,
$\tau_R\rightarrow \tau_R$; $\mu_R\rightarrow \omega^2 \mu_R$ and
$e_R\rightarrow \omega e_R$. If we have a Higgs doublet $H$ which is invariant
under this $Z_3$ symmetry, then the following Yukawa coupling between
fermions and the Higgs bosons is both gauge and $Z_3$ invariant:
$$L_1=h_0\bar{\Psi}_{0L}H\tau_R +h_1 \bar{\Psi}_{1L}H\mu_R +h_2\bar{\Psi}_{2L}
H e_R+~h.c.\eqno(10)$$
where
$\Psi_{0L}= {{1}\over{\sqrt{3}}}({\psi}_{1L}+{\psi}_{2L}+{\psi}_{3L})$;
$\Psi_{1L}={{1}\over{\sqrt{3}}}(\psi_{1L}+\omega\psi_{2L}+\omega^2\psi_{3L})$
and $\Psi_{2L}={{1}\over{\sqrt{3}}}(\psi_{1L}+\omega^2\psi_{2L}+
\omega\psi_{3L})$. We add to this theory
 a Higgs triplet $\Delta_L$ with lepton number 2
which has a small vev which can be generated by see-saw mechanism for
vev's {\it a la} ref.\cite{mohsen} by including a gauge singlet complex higgs
$\Delta_R$ which has a large vev and is responsible for the right-handed
Majorana neutrino mass in the see-saw matrix of eq.3. If the $\Delta_L$
is assumed to be a $Z_3$ singlet, then there are three gauge and $Z_3$
invariant couplings  possible in the model: $\psi^T_L\psi_L\Delta_L$,
$\psi^T_LS\psi_L\Delta_L$ and $\psi^T_LS^2\psi_L\Delta_L$ where we
have assumed that $\psi^T_L=(\psi_{1L},\psi_{2L},\psi_{3L})$ is a row
vector. In order to have a degenerate scenario, we will keep only the
first invariant as part of our Yukawa Lagrangian :
$$L_2~=~f_{LL}\psi^T_L\psi_L\Delta_L~+~h.c.\;\eqno(11)$$

 As indicated after
the $\Delta_L$ has a vev, it will generate a common Majorana mass for the
light neutrinos. Finally, we will include in the theory three more Higgs
doublets, which we denote by $\phi_i$ with $i=1,2,3$ such that they transform
under the the $Z_3$ symmetry exactly as three lepton doublets. If we
further assume the right handed neutrinos to be singlets under the $Z_3$,
then there are several couplings allowed but we keep only a subset of those
terms:
$$L_3= \bar{\psi}_{1L}\phi_1 \nu_{1R}+\bar{\psi}_{2L}\phi_2\nu_{2R}
+\bar{\psi}_{3L}\phi_3\nu_{3R}+\Sigma_i(f_i\nu_{iR}\nu_{iR}\Delta_R)
+~h.~c.  \eqno(12)$$.

After symmetry breaking $\langle H^0\rangle\neq 0$, $\langle\phi^0_i\rangle
\neq 0 $ as well as vev's for $\Delta^0_{L,R}$, one has a see-saw matrix
for the neutrino sector, where there is no generation mixing and one has
the modified see-saw matrix of eq.3 that leads to eq.4 for the $\nu_{i}$
masses. The mixings arise purely from the charged lepton sector and it is
easily seen that it has the desired maximal form as in eq.(5). We wish to
emphasize again that we have not provided a technically natural derivation
of the maximal mixing in the strict field theoretical sense but shown
that any theory for such a mass and mixing pattern ought to have the
generational $Z_3$ symmetry given in eq.(9). We envision this as arising
from some high scale horizontal $SO(3)$ or
$U(3)$ symmetry which after breakdown
leaves the above $Z_3$ symmetry at low energies.

In conclusion, we have shown how a maximal CP-violating mixing among the
three  or two light Majorana neutrinos has the potential to accomodate the hot
dark matter neutrino  and the
$\beta\beta_{0\nu}$ data without at the same time
contradicting the neutrino vacuum oscillation
solutions to the solar and atmospheric neutrino data. While a detailed
fit to the solar neutrino data discussed by other authors for similar
scenarios is not completely satisfactory, at the present state of
things, we are not discouraged by it. Clearly as the experimental
situation is further sharpened, the fate of these models will be completely
decided. We have also attempted to derive the maximal mixing pattern
in extensions of the standard model.


\begin{thebibliography}{99}

\bibitem{seesaw} M. Gell-Mann, P. Ramond and R. Slansky, {\it Supergravity},
ed. D. Freedman et al, (North Holland, 1979); T. Yanagida, Proceedings of
the conference of the baryon Number of the Universe, ed.   (1979); R. N.
Mohapatra and G. Senjanovi\'c, {\it Phys. Rev. Lett.} {\bf 44}, 912 (1980).

\bibitem{chang} D. Chang, R. N. Mohapatra, J. Gipson, R. E. Marshak and
M. K. Parida, {\it Phys. Rev. } {\bf D31}, 1718 (1985); N. G. Deshpande,
E. Keith and P. Pal, {\it Phys. Rev.} {\bf D46}, 2261 (1992).

\bibitem{ch} CHORUS collaboration preprint CERN-SPSC/90-42 (1992);
NOMAD collaboration preprint CERN-SPLSC/91-21 (1992).

\bibitem{cald} D. Caldwell and R. N. Mohapatra, {\it Phys. Rev.} {\bf D48},
3259 (1993) and {\it ibid} {\bf D50}, 3477 (1994).

\bibitem{gaisser} W. Frati et al. {\it Phys. Rev.} {\bf D 48}, 1140 (1993).

\bibitem{mohsen} R. N. Mohapatra and G. Senjanovi\'c, {\it Phys. Rev. Lett.}
{\bf 44}, 912 (1980); {\it Phys. Rev.} {\bf D23}, 165 (1981).

\bibitem{many} D. Caldwell and R. N. Mohapatra, ref\cite{cald};
D. G. Lee and R. N. Mohapatra, {\it Phys. Lett.} {\bf B 329}, 463 (1994);
S. T. Petcov and A. Smirnov, SISSA preprint 113/93/EP (1993);
P. Bamert and C. Burgess, {\it Phys. Lett.B} ( to appear) (1994);
A. Joshipura, PRL preprint (1993);
A. Ioanyssian and J. W. F. Valle, {\it Phys. Lett.} {\bf B 332}, 93 (1994);
C. Albright and S. Nandi, Fermilab Preprint (1994);
K. S. Babu and S. Pakvasa, {\it Phys. Lett.} {\bf B172}, 360 (1986).

\bibitem{klap} For recent reviews, see H. Klapdor-Kleingrothaus,
{\it Prog. in Part. and Nucl. Physics} {\bf 32}, 261 (1994);
 D. Caldwell et al., {\it Nucl. Phys.} (Proc. Supp.)
{\bf B13}, 547, (1990); M. Moe and P. Vogel, {\it Ann. Rev. of Nucl. and
Part. Sc.} ( to appear) (1994).

\bibitem{klap2} H. V. Klapdor-Kleingrothaus et al, {\it Proceedings of
the 27th International Conference on High Energy Physics}, Glasgow (1994)
(to appear); IGEX collaboration also reports an upper limit on $m_{\nu_e}$
of order 1 eV ( private communication to S. Nussinov).

\bibitem{nuss} S. Nussinov, {\it Phys. Lett.} {\bf 63B}, 201 (1976).

\bibitem{wolf} L. Wolfenstein, {\it Phys. Rev.} {\bf D 18}, 958 (1978).

\bibitem{nemo} {\bf NEMO 3} proposal, LAL 94-29 (1994).

\bibitem{krastev} P. Krastev and A. Smirnov, {\it Phys. Lett.B} ( to appear).
N. Hata and P. Langacker, Pennsylvania preprint (1994).

\bibitem{solar} R. Davis, {\it Proceedings of the $6^{th}$ International
Conference on Neutrino Telescopes}, ed. M. Baldo-ceolin (to appear) (1994);
Y. Suzuki, {\it ibid.}; Gallex Collaboration, {\it Phys. Lett.} {\bf B 327},
 377 ( 1994); SAGE Collaboration, in {\it Proceedings of the $27^{th}$
International Conference on High Energy Physics}, Glasgow, July (1994);
 For reviews, see T. Kirsten, {\it Proceedings of the
International Conference on Non-Accelerator Physics, Bangalore, India},
ed. R. Cowsik et al. ( to appear) (1994); T. Bowles, {\it ibid.}

\bibitem{kim}   C. W. Kim and J. A. Lee, Johns Hopkins preprint, (1993);
C. Giunti, C. W. Kim and J. D. Kim, SNUTH-94-109 (1994).

\bibitem{bere} P. Krastev and S. T. Petcov,{\it Phys. Rev. Lett.} {\bf 72},
 1960 (1994); N. Hata, UPR-0605T, (1994); Z. Berezhiani and A. Rossi,
INFN FE-11-94.

\bibitem{louis} W. Louis, in {\it XVI Conf. on Neutrino Phys. and Astrophys.},
Eilat, Israel (1994).

\end{thebibliography}
\end{document}